%% file: main.tex
\documentclass[conference]{IEEEtran}
\usepackage[utf8]{inputenc}
\IEEEoverridecommandlockouts
\usepackage{cite}
\usepackage{amsmath,amssymb,amsfonts}
\usepackage[pdftex]{graphicx}
\usepackage{textcomp}
\usepackage{float}
\usepackage[dvipsnames]{xcolor}
\usepackage[acronym,nomain,nonumberlist,nogroupskip]{glossaries}
\usepackage[caption=false,font=normalsize,labelfont=sf,textfont=sf]{subfig}
\usepackage{tabularx}
\usepackage{longtable}
\usepackage{multirow}
\usepackage{multicol}
\usepackage[bookmarks=false,hypertexnames=true]{hyperref} 
\usepackage{booktabs}

\title{Dynamic Reliability: Reliably Sending Unreliable Data}
\IEEEaftertitletext{\vspace{-2\baselineskip}}
\author{
\IEEEauthorblockN{Omar~Nassef, Federico~Chiariotti, Stephen~Johnson, Toktam~Mahmoodi}
\thanks{Omar Nassef (omar.nassef@kcl.ac.uk) and Toktam Mahmoodi (toktam.mahmoodi@kcl.ac.uk) are with the Centre of Telecommunication Research, Kings College London, UK. Federico Chiariotti (chiariot@dei.unipd.it) is with the Department of Information Engineering, University of Padova, Italy, and with the Department of Electronic Systems, Aalborg University, Denmark. Stephen Johnson (stephen.h.johnson@bt.com) is with British Telecommunications plc, UK. This project is jointly funded by BT, the UK Engineering and Physical Sciences Research Council (EPSRC), and King's College London Centre for Doctoral Studies (CDS).
}
}

\begin{document}

\setlength{\intextsep}{1mm}
\setlength{\textfloatsep}{1mm}

\input{acrons}

\maketitle

\begin{abstract}

    5G and Beyond networks promise low-latency support for applications that need to deliver mission-critical data with strict deadlines. However, innovations on the physical and medium access layers are not sufficient. Additional considerations are needed to support applications under different network topologies, and while network setting and data paths change. Such support could be developed at the transport layer, ensuring end-to-end latency in a dynamic network and connectivity environment. In this paper, we present a partial reliability framework, which governs per-packet reliability through bespoke policies at the transport layer. The framework follows a \textit{no-ack} and \textit{no-retransmit} philosophy for unreliable transmission of packets, yet maintains cooperation with its reliable counterpart for arbitrary use of either transmission mode. This can then address latency and reliability fluctuations in a changing network environment, by smartly altering packet reliability. Our evaluations are conducted using mininet to simulate real-world network characteristics, while using a video streaming application as a real-time use-case. The results demonstrate the reduction of session packet volume and backlogged packets, with little to no effect on the freshness of the packet updates.

\end{abstract}


\input{introduction}
\input{soa}
\input{methodology}
\input{results}
\input{conclusion}
\bibliographystyle{IEEEtran}
\bibliography{references}

\end{document}

%% file: acrons.tex
\newacronym{mptcp}{MPTCP}{Multipath TCP}
\newacronym{mpquic}{MPQUIC}{Multipath QUIC}
\newacronym{tcp}{TCP}{Transmission Control Protocol}
\newacronym{hol}{HoL}{Head of Line Blocking}
\newacronym{ietf}{IETF}{Internet Engineering Task Force}
\newacronym{emb}{eMBB}{Enhanced Mobile Broadband}
\newacronym{mtc}{mMTC}{Massive Machine Type Communication}
\newacronym{urllc}{URLLC}{Ultra-Reliable Low-Latency Communication}
\newacronym{iot}{IoT}{Internet of Things}
\newacronym{mmwave}{mmWave}{Millimeter Wave}
\newacronym{sctp}{SCTP}{Stream Control Transmission Protocol}
\newacronym{dccp}{DCCP}{Datagram Congestion Control Protocol}
\newacronym{tcp}{TCP}{Transmission Control Protocol}
\newacronym{udp}{UDP}{User Datagram Protocol}
\newacronym{rtt}{RTT}{Round-Trip Time}
\newacronym{srtt}{SRTT}{Smoothed-RTT}
\newacronym{aoi}{AoI}{Age of Information}
\newacronym{qos}{QoS}{Quality of Service}
\newacronym{ack}{ACK}{Acknowledgement}
\newacronym{kpi}{KPI}{Key Performance Indicator}
\newacronym{ran}{RAN}{Radio Access Network}
\newacronym{soa}{SOA}{State of the Art}

%% file: introduction.tex
\section{Introduction}
\label{intoduction}

The roll-out of 5G has enabled a plethora of use cases and new applications, which generally fall under three broad categories: \gls{emb}, \gls{mtc} and \gls{urllc} \cite{5g_pillars}. Most human-generated traffic is \gls{emb}, which supports higher data rates, mobility and dense connectivity thanks to the capability of connecting to micro- and nano-cells. Other scenarios that can be characterized as \gls{emb} include large-scale distributed machine learning~\cite{dl}, which is expected to generate massive volumes of traffic. On the other hand, \gls{mtc} focuses on high density \gls{iot} device deployments, and specifically aims at supporting longer transmission distances and reducing energy consumption by adopting lower data rates~\cite{nb-iot-2}. Finally, \gls{urllc} includes scenarios that require near perfect reliability (i.e., an error probability below $10^{-5}$) with a maximum latency of $1$~ms. Its use cases extend to cooperation between autonomous vehicles~\cite{vehicles} and the Tactile Internet~\cite{haptic}.

The research on adapting the \gls{ran} to the new requirements of these different classes of traffic is extensive, but the \gls{ran} is not the only factor affecting quality of application's delivery. The presence of cross-traffic can cause significant delays and congestion, and even without it, self-queuing delay is a well-known problem in wireless networks, as probing the connection capacity aggressively can increase the latency~\cite{polese2019survey}~\cite{tcp-asymetry}.

Additionally, different applications and use-cases expect different levels or latency and/or reliability. This can make optimization even more complex: meeting different requirements for several applications, all with independent adaptation mechanisms, is a significant challenge. Even the nature of the metrics themselves might be different: one recent example is \gls{aoi}, which has currently risen in prominence for several \gls{iot} applications~\cite{yates2021age}, as it can measure the freshness of the data available to the receiver, incorporating the data generation process as well as network-related aspects.

An important detail in \gls{aoi} optimization is that not every packet is needed: if new updates are frequent enough, the more recent data supersedes the older information, and a certain level of packet loss is acceptable. This is opposed to the extreme reliability requirements of \gls{urllc} traffic, but there are several use cases and scenarios where it intuitively makes sense, e.g., cooperative autonomous driving: if a car frequently sends information about its position and direction, the loss of any single sensor reading can be recovered from~\cite{quicest}, keeping in mind that urgent safety messages will be sent as \gls{urllc} traffic, ensuring that mission-critical information, such as collision warnings, is reliably transmitted.

These real-time, relatively loss-tolerant information flows may not require end-to-end reliability mechanisms, as retransmissions can increase delay and traffic on the network: in order to better support the concurrent \gls{urllc} and \gls{emb} flows, which need network resources due to their stricter requirements, these applications can send data unreliably. In this paper, we introduce dynamic reliability: a novel idea for partial reliability. Dynamic reliability permits assignment of per-packet reliability status at the transport layer, governed by tailored policies, opposed to the traditional application layer manner. The framework follows a true unreliable transmission philosophy of removing \glspl{ack} and retransmissions. The proposed solution is a modification of the QUIC protocol~\cite{quic}, which can support concurrent flows with different reliability requirements, dynamically setting the reliability of each packet while considering the state of the connection.

The proposed solution is examined in detail with respect to a number of network topologies and links, considering metrics such as session packet volume, backlogged packets and \gls{aoi}, to ascertain the feasibility of dynamic reliability in real-time scenarios. Furthermore, we explore a number of governing reliability policies to show the adaptability of dynamic reliability to different requirements. The contributions of this paper include:
\begin{itemize}
    \item A dynamic reliability framework that allows per-packet assignment of reliability at the transport layer, depending on the governing reliability policies.
    \item Performance evaluation of a number of reliability policies for dynamic reliability. The policies explore different logic and complexity for reliability assignment based on path and network measurements. 
    \item A comparison between the different dynamic reliability logic over wireless communication using different frequencies, including \gls{mmwave}, sub-6GHz and Wi-Fi.
\end{itemize}

The remainder of this paper is structured as follows: Sec. \ref{soa} showcases the \gls{soa}, informing the reader on QUIC and the different implementations of partial reliability. Sec. \ref{methodology} delves into the system model and design of dynamic reliability, while Sec. \ref{results} analyses the result of implementing dynamic reliability in simulated network environments. Lastly, the conclusion and future works are discussed in Sec. \ref{conclusion}.

%% file: soa.tex
\section{State of the Art}
\label{soa}

Although the transport layer does not consider any \gls{ran} parameters directly, indirect end-to-end measurements such as \gls{rtt} are used to assess the bandwidth, latency and loss rate. QUIC is one protocol of many that utilises these path properties for end-to-end communication services.
QUIC was designed by Google to circumvent latency issues that were associated with the traditional \gls{tcp}. More specifically, it allowed multiplexing a session into different streams between the two endpoints, to bypass \gls{hol}. Each stream in QUIC is given a Connection ID and treated as a separate flow, this allows each stream to terminate or migrate without affecting the rest of the session. Each stream also has its own packet sequence independently from other streams, maintaining its own in-order delivery system.

Despite the fact that QUIC runs over \gls{udp}, its transmission is reliable and in-order for each stream. That being said, recently a \emph{Datagram Extension} has been added to QUIC to enable unreliable transmission of packets as well \cite{datagram}. A reliable and unreliable stream between the same end-points can share a single handshake and proceed with transmission normally, but it is left for the application layer to differentiate between datagram flows, burdening the application layer with a larger complexity as well as offering no re-ordering for the datagram packets. Even though datagram frames are unreliable, they are still ACK-eliciting and their rate is affected by the congestion window.

In the most general definition, partial reliability allows for a transport protocol to send reliable and unreliable data when the application deems fit. While the traditional fully reliable approach has been challenged by transport protocols such as \gls{sctp}~\cite{pr_sctp}, \gls{dccp}~\cite{pr-dccp} and even \gls{tcp} \cite{pr_tcp}, there is no unified concept of partial reliability, or a shared technical solution. Each protocol implements their distinct definition of partial reliability with disparate design aims. The QUIC datagram extension also goes in this direction, but the decision on which data should be sent reliably is still an open research question.

QUICSilver \cite{quicsilver} uses predicted deadline awareness to guide the decision making for sending frames reliably. The aim of QUICSilver is to reduce the occurrence of stalls in video transmission, which it has managed to achieve compared to vanilla QUIC, albeit only to only low-quality video transmissions. As stated by the author, the freshness checks to determine the staleness of the frames introduce a non-negligible delay which inherently affects the playback of the video transmission.

Furthermore, ClipStream \cite{clipstream} sends Intra-Frames (i.e., frames that contain sufficient data to display the whole image) and end-of-stream markers reliably, while Predicted Frames (i.e., differentially encoded frames based on the previous I-frame) are sent unreliably. Opportunistic retransmission is used, sending new data instead of retransmissions for the unreliable streams. That being said, this work does not take into account the strict playback delays when considering the retransmission of reliable frames.

Alternatively, QUIC-EST \cite{quicest} adopts a multi-sensory use case as their variation of a real-time application. The aim is to reduce the probability of undelivered frames in blocking fresh content transmission. The design goal is to have every object sent on its own stream with a priority allocation based on the value of information that frame induces. When the packets on a single stream are acknowledged, the stream can be reused for consecutive new objects. If a packet is lost or excessively delayed, the whole stream is discarded, opting out of retransmissions. Although the approach can be generalised to many mission-critical and real-time applications, it requires the application layer to assign a priority value to objects.

In this work, we consider a new approach for partial reliability. We focus on per-packet partial reliability at the transport layer piloted by policies, which emphasises reactivity to fluctuations in network measurements, rather than the traditional method of partial reliability assignment at the application layer. We remove the need for ack-elicitation following a true unreliable philosophy, which has not been explored in the literature. Additionally, we incorporate ``timed reliability'' from \gls{sctp} \cite{pr_sctp}, but utilise an occupancy-driven metric rather than time-based: when the buffer reaches a threshold of packets received, all the packets that have not been sent to the application layer are discarded.

%% file: methodology.tex
\section{Architecture and System Model}
\label{methodology}

Although QUIC has already standardised an extension for unreliable packet sending, the protocol requires \glspl{ack} to ensure correct functionality. While \gls{ack} packets assist in loss recovery and congestion control, they defy the true meaning of unreliable transmission. Furthermore, the receiver is encouraged to delay sending back \gls{ack} frames, which may lead to unpredictable behaviour. Finally, the co-existing nature of reliable and unreliable frames is non-existent at the transport layer, relegating all the responsibility and complexity to the application layer. This behaviour can be sub-optimal, especially for real-time and \gls{aoi}-oriented applications.

Therefore, the main novelty of this work is to introduce the concept of dynamic reliability, which seamlessly enables and disables reliable packet sending through tailored policies, achieving reduced traffic load when needed to prioritise transmission of fresh packets. The decision to transmit reliably or unreliably could depend on the characteristics of network and applications. Unreliable packets do not elicit an \gls{ack} and co-exist with its reliable counter-part at the transport layer. For example, switching to unreliable mode will be attractive for real-time, loss-tolerant applications that could yet still benefit from end-to-end stream multiplexing, security, and other features of QUIC, as compared to plain UDP.

\begin{figure}[t]
    \centering
    \includegraphics[width=0.48\textwidth]{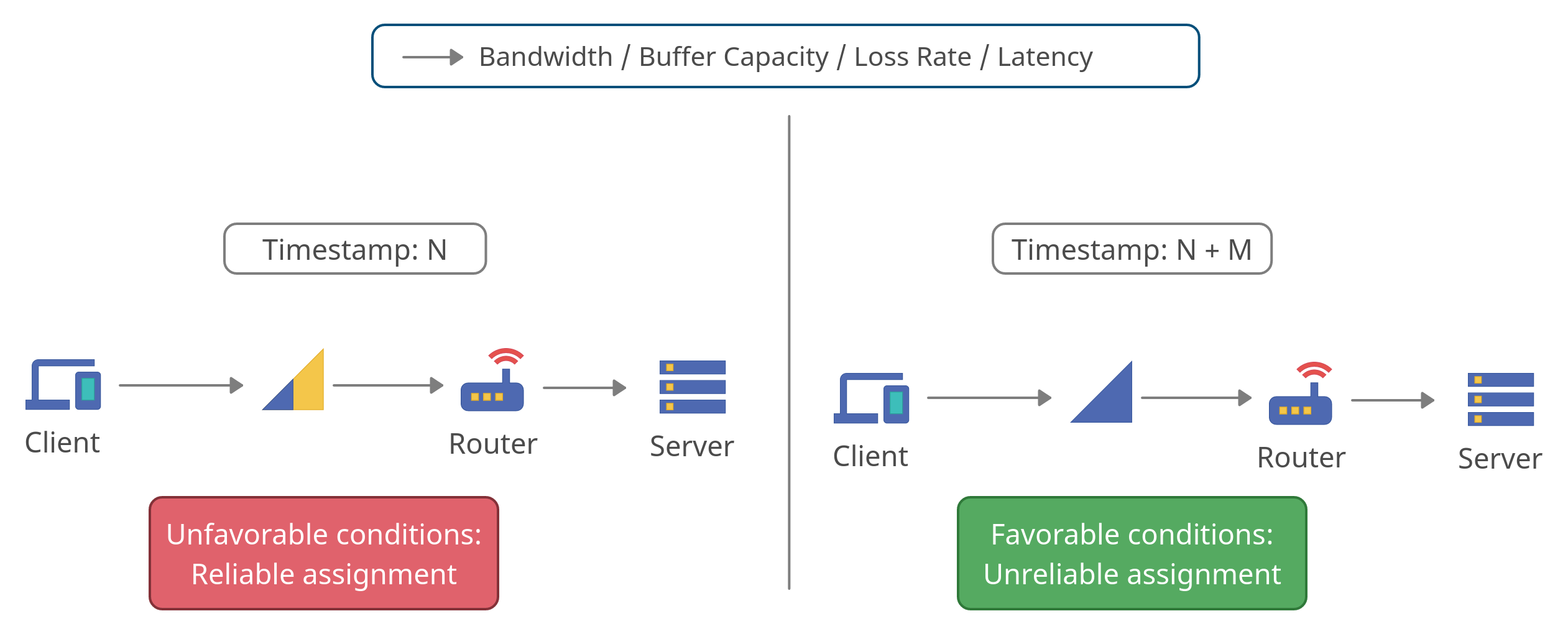}
    \caption{Illustration of Dynamic Reliability logic in a mobile network architecture. As the quality of the network improves, unreliable transmission is adopted. The favorability is determined by the reliability policy.}
    \label{fig:sim_architecture}
\end{figure}

We examine the performance of the proposed dynamic reliability framework using Mininet, and simulate both the end-to-end and intermediate nodes, as well as, the real-world network characteristics, in a setup depicted in Fig. \ref{fig:sim_architecture}. In this setup, the client live streams an HD video ($\approx 60$ frame-per-second) to the server with a variable send-rate dependant on the live-stream and network conditions. Furthermore, the links connecting the nodes are selected from Wi-Fi, mobile Sub-6GHz and \gls{mmwave} with properties described in Table \ref{tab:path_char}. To isolate the performance gain of dynamic reliability in the simulation, we negate the presence of other services, and reduce the load of the network to only the essential components of the live stream. We also consider bursty loss scenarios, modeling the connection as a Gilbert-Elliott two-state channel \cite{gilbert_elliot}. 

\begin{table}[b]
    \caption{Path characteristics for different link technologies \cite{mmwave_bandwidth,mmwave_measurements,path_characteristics}}
    \centering
    \begin{tabular}{c|ccc}
        \toprule
         \multirow{2}{*}{\textbf{Parameter}} & \multicolumn{3}{c}{\textbf{Technology}}  \\
         & Sub-6GHz & Wi-Fi & mmWave (LoS) \\
         \midrule
         Capacity (Mb/s) & 1100 & 30 & 2500\\
         Delay (ms) & $27.4 \pm ~6.4$  & $20 \pm ~10$ & $2 \pm ~1$  \\
         Loss Ratio (Percent) & 0.1 & 0.7 & 0.1 \\ 
         \bottomrule
    \end{tabular}
    \label{tab:path_char}
\end{table}

\subsection{Dynamic Reliability}

The addition of extra \gls{ack} packets can increase congestion during traffic bursts, reducing the maximum load that a network can support. This has been observed, e.g. in WiFi connections, in which even short \gls{tcp} \gls{ack} packets on the uplink can significantly reduce downlink throughput~\cite{zubeldia2012averting} due to the contention for access to the channel. Furthermore, real-time, \gls{aoi}-oriented applications rarely require retransmission of stale frames, opting instead for the latest frames as they relatively hold much more value. This is a common assumption in the \gls{aoi} literature, and intuitively makes sense when we consider the immediate use of sensor readings~\cite{yates2021age}. 

To achieve dynamic reliability, we extend QUIC by adding a new frame type, which resembles the reliable frame, albeit a different frame type byte. This new frame flags \texttt{no-ack} and \texttt{no-retransmit} to represent an unreliable packet. It is important to note that the packet sequencing remains the same with unreliable and reliable flows, the only difference being that unreliable packets do no solicit an acknowledgement, and hence, are never retransmitted.

However, as most congestion control mechanisms use \glspl{ack} to infer congestion and measure capacity, the lack of acknowledgements from the unreliable segment of transmission means that the congestion window may not be strictly accurate. Despite that, the unreliable transmission will still follow the rate presented by the congestion window, which places emphasis on fairness of the protocol with other traffic on the link. At the same time, adhering to a stale congestion window is sub-optimal, and the reliable segment might not be enough to get an accurate picture of the channel. We then need to introduce probing to update the congestion window and path measurements such as \gls{rtt}, preventing the use of stale feedback in the transmission of packets.

The features of the proposed dynamic reliability framework are as follows:
\begin{itemize}
    \item Permitting of per-packet assignment reliability through policies located at the transport layer.
    \item Ability to send reliable and unreliable packets concurrently along the same path, session or stream. This provides further granular control over the reliability of parallel, multiplexed and distinct traffic type transmissions an application may adopt through streams;
    \item No requirement for separate packet sequences between unreliable and reliable packets. This is the main enabler for unreliable and reliable packets to co-exist in the same session;
    \item Unreliable packets do not elicit \glspl{ack}, therefore, less network resources are consumed;
    \item Loose coupling of congestion control and loss recovery mechanisms; both \gls{ack} and non-\gls{ack} based mechanisms can be integrated.
    \item Maintained inter- and intra- flow fairness by adhering to the congestion control. 
\end{itemize}

\begin{figure*}[ht]
    \centering
    \includegraphics[scale=0.15,trim={0 2.5cm 0 6cm},clip]{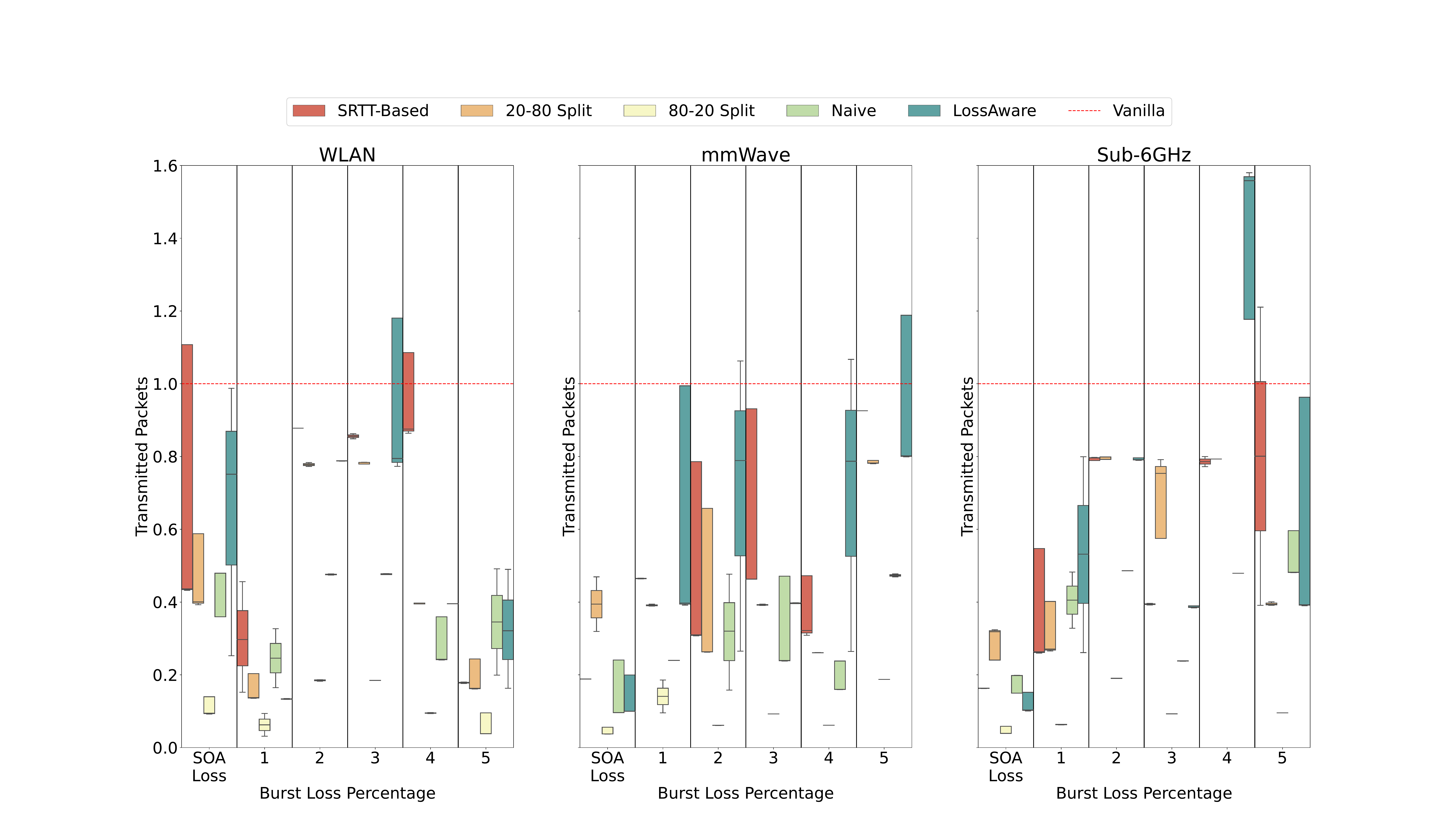}
    \caption{Normalised frequency of transmitted packets with burst and \gls{soa} loss values against the dynamic reliability logic for different network topologies.}\vspace{-0.4cm}
    \label{fig:sent_burst}
\end{figure*}

\begin{figure*}[ht]
    \centering
    \includegraphics[scale=0.15,trim={0 2.5cm 0 6cm},clip]{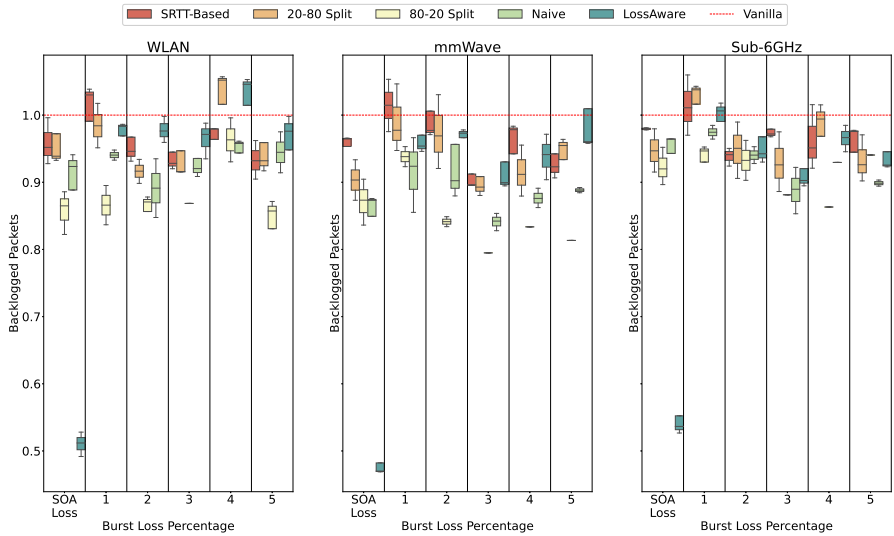}
    \caption{Normalised frequency of backlogged packets with burst and \gls{soa} loss values against the dynamic reliability logic for different network topologies.}
    \label{fig:backlog_burst}\vspace{-0.4cm}
\end{figure*}

\subsection{Reliability Policies}

The reliability policies can be tailored to fit the application, network or holistic constraints. Though, it is important to note that the policies perform at the transport layer and cannot be altered at the application level. The verdict of the policy is used to set the type byte of the frame to indicate its reliability status. Since the policy is ingrained in the QUIC protocol, it has access to a diversity of information available from the transport protocol e.g. \gls{rtt}, congestion window and bytes in-flight to name a few.

A number of reliability policies are explored in this work with varying levels of complexity and intelligence: the Naive, 20-80 split and 80-20 split policies use a static unreliability packet ratio, providing a performance baseline. On the other hand, the \gls{srtt}-based and loss-aware policies take into account the state of the network to guide the packet reliability assignment thus providing a smarter reliability policy.  We take the vanilla implementation of QUIC to be the benchmark for the dynamic reliability implementation.

Our first scenario follows a \textit{naive} approach in transmitting packets, with the purpose of highlighting its inadequacy in obtaining greater performance optimisation compared to more complex policies.
Following its name, the naive policy randomly flags $50\%$ of packets as reliable. The policies for subsequent scenarios are detailed below:

\subsubsection{80-20 Split} Similar to the Naive method, packets have an 80\% probability of being flagged as reliable.

\subsubsection{20-80 Split} The polar opposite of 80-20, where packets have a chance of being flagged as reliable 20\% of the time.

\subsubsection{SRTT-Based Logic}

Utilises the \gls{srtt} as a basis for appraising the network condition. The \gls{srtt} is calculated from reliable packets, adhering to the method used by QUIC to track the \gls{rtt} and its derivatives. If the latest \gls{rtt} is lower than the \gls{srtt}, it can be assumed that the network conditions are acceptable for marking packets as unreliable.

\subsubsection{Loss-Aware Logic}

Curated for bursty loss scenarios, this policy takes into account the exponentially weighted moving average \cite{ewma} of the session loss rate. The loss rate is shown in Eq. \ref{la_decision_making}, while Eq. \ref{smoothing-eq} adopts the exponential smoothing. Let $P_{s}$ be the packets that are sent, $P_{us}$ be the packets are unreliably sent and $P_{r}$ be the packets received which are not \gls{ack}-eliciting. The measurements are taken at $i = 1$, which serves as the start of the session.

\begin{equation}
\label{la_decision_making}
\lambda = \frac{\sum_{i=1}^{n-1} P_{s} - \sum_{i=1}^{n-1} P_{us}}{\sum_{i=1}^{n-1} P_{r}}
\end{equation}

\begin{equation}
\label{smoothing-eq}
 \omega_{t} = \alpha \lambda_{t-1} + (1 - \alpha) \omega_{t-1},
\end{equation}

where $n$ denotes the discrete sending times. $\alpha$ is the weighting constant $ 0 \leq \alpha \leq 1$, which prescribes the importance of the previous measurements. In a bursty scenario, the measurements change frequently, and past recordings hold little relative value. However, in order to detect a burst, some past measurements are required: as such, we set the discount value to 0.8. The real-time constraint $RT$ is set at 5\%, which is the maximum tolerable loss rate for a real-time application, where $\omega_{t} \leq RT$ permits unreliable sending.

%% file: results.tex
\section{Results}
\label{results}

\begin{figure*}[ht]
    \centering
    \includegraphics[scale=0.15,trim={0 2.5cm 0 5cm},clip]{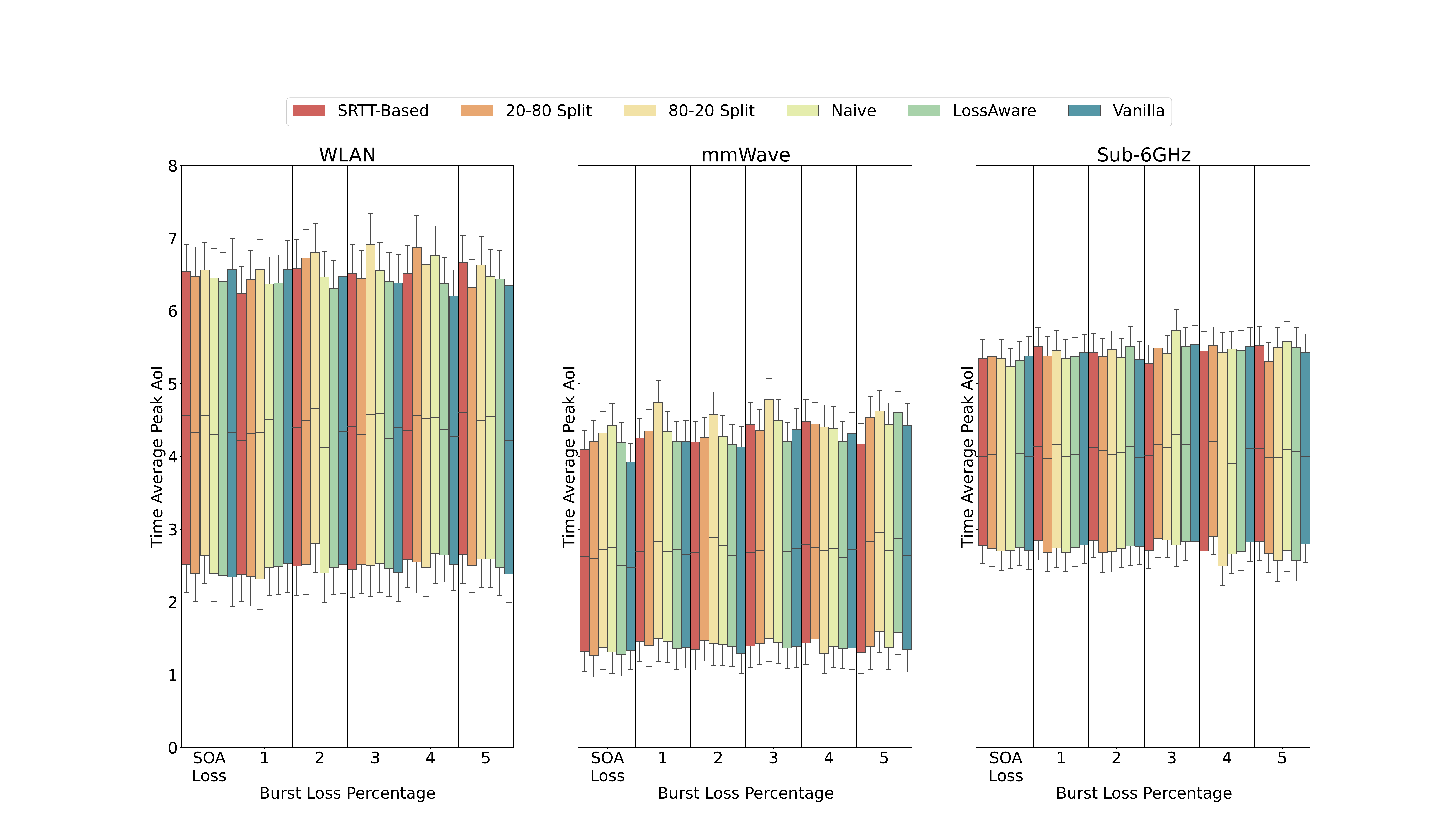}
    \caption{The time average peak Age of Information with burst and \gls{soa} loss values against the dynamic reliability logic for different network topologies.}
    \label{fig:aoi_burst}\vspace{-0.4cm}
\end{figure*}

This paper focuses on both transport layer and application layer metrics to determine the feasibility of dynamic reliability. For this, we have selected the session packet volume, as transmitted, retransmitted, lost and backlogged packets as \glspl{kpi} for the transport layer; while focusing on the \gls{aoi} for the application layer. The \gls{aoi} was chosen as a crucial indicator for the freshness of packets in real-time applications. More specifically, this work adopts the time average peak \gls{aoi} equation \cite{aoi_equation} depicted in Eq. \ref{aoi}, where $\Delta(r_{i+1})$ is the $i$th update at the time it was received at the server, for a session time period of $\tau$.

\begin{equation}
    \label{aoi}
    \gls{aoi}_\tau = \frac{1}{n-1}\sum_{i=1}^{n-1} \Delta(r_{i+1})
\end{equation}

We include a comparison between the vanilla QUIC implementation which does not enjoy the dynamic reliability extension, with a number of dynamic reliability policies. The tests were run a number of times for statistical significance, with the mean value of vanilla implementation used as a baseline for comparison. The topology utilised both random loss and bursty loss to explore the bounds of dynamic reliability. The \gls{soa} loss in the figures correspond to the loss values presented in Table. \ref{tab:path_char}, for ease of comparison between bursty and random loss scenarios.

\subsection{Transport-Layer KPIs}

To analyse the performance gain at the transport layer due to dynamic reliability, the volume of transmitted and backlogged packets is examined. The figures are in the form of boxplots, which take the vanilla implementation as a benchmark, depicted as the red dashed line.

As seen in Fig. \ref{fig:sent_burst}, the loss plays a crucial role in the performance of the reliability policies. The policies under random loss did incredibly well for the networks with a larger capacity, namely \gls{mmwave} and Sub-6~GHz, whereas for burst loss, the lower network capacities had a larger packet reduction. With the increase in burst loss, the behaviour of the set split reliable policies became unpredictable, if a reliable assignment happened to coincide with a burst loss, the number of transmitted packets increases, and vice versa. On the other hand, in smarter policies, such as Loss-Aware, the performance lightly matched the vanilla baseline, as the reliable assignment dominated the session to compensate for a higher burst loss. Not only that but, the burst loss also impacted the variance of the transmitted packets for the policies.

Unsurprisingly, the unreliable focused policy, 80-20 split, outperformed other policies for all topologies in random and bursty loss scenarios, with an approximate reduction of 80\%. That being said, the majority of the policies reduced the transmitted packets on the link by approximately 70\% for random loss, while the reduction started at $\approx 15\%$ and decreased as the loss increased for the burst loss scenario.

The retransmitted and lost packets, not shown due to space limitations, followed the same trend as the transmitted packets for the random loss scenarios. However, for the burst loss scenarios, the larger capacity networks had a lower reduction in the retransmitted and lost packets. This can be seen as a favorable outcome since the lower capacity networks are scarce on resources. It is important to note that the Loss-Aware policy mimicked the vanilla approach as the burst loss increased, signifying the overwhelming appointment of reliable packets in adapting to the harsh burst loss conditions.
 
Alternatively, Fig. \ref{fig:backlog_burst} clearly shows a stark comparison between the policies and loss scenario in the reduction of the backlogged packets. The Loss-Aware policy for random loss scenario reduced the backlogged packets by up to 50\%, beating all other policies by approximately 30\%. Furthermore, it is clear that the unreliability focused policies resulted in the lowest backlog for the session. In comparison, we notice that the burst loss and the backlogged frequency have a positive correlation, where the maximum reduction of the backlogged packets for the policies is at most 20\%. Much like the transmitted packets, the probability of a burst loss occurrence plays a vital role in the number of retransmissions sent and by extension the number of backlogged packets. Thus, we can conclude that the stress placed on the buffer is a result of the reliable packets which is tightly coupled with the congestion on the session. Whereas, unreliable focused policies did not encounter such a phenomenon regardless if it was experiencing a burst loss.

\subsection{Application-Layer KPIs}

The feasibility of dynamic reliability for real-time applications can be determined by the \gls{aoi}, with comparison across different topologies and policies. If we take a strict approach and consider anything below $10$~ms is real-time \cite{real-time}, then all the reliability policies passed that requirement, which is attractive for real-time applications, as shown in Fig. \ref{fig:aoi_burst}. Utilising the median as an estimate of the runs, the policies in the WLAN and Sub-6~GHz topology with random loss floated around $4-5$~ms with negligible difference, while the \gls{aoi} for \gls{mmwave} was $\approx 2-3$~ms. It is clear that the \gls{aoi} and the network capacity have a negative correlation, as the network capacity decreases, the \gls{aoi} increases. The same correlation is extended to the bursty loss scenarios, where \gls{mmwave} dominated the other topologies. That being said, it is crucial to note that the \gls{aoi} for the reliability policies is often slightly better than or equal to the \gls{aoi} of the vanilla implementation, proving that dynamic reliability reduces the congestion of the session at no cost to the \gls{aoi}.

%% file: conclusion.tex
\section{Conclusion}
\label{conclusion}

Dynamic reliability can be used to react and cater to varying network topolgies with fluctuating measurements. Adhering to governing policies in changing the per-packet reliability status to deal with bandwidth, loss and delay constraints without deteriorating the \gls{aoi}. Less resource intensive requirements can reduce the cost of communication for operators and free scarce resources for alternative traffic.

Future work will consider a holistic dynamic reliability policy for interchangeable network conditions whilst focusing on the over all quality of service from both a user and an operator perspective.